\documentstyle[12pt,epsfig]{article}
\begin{document}

\begin{titlepage}
\rightline{April 2008}
\vskip 2cm
\centerline{\Large \bf  
Mirror dark matter and the new DAMA/LIBRA results:} 
\vskip 0.5cm
\centerline{\Large \bf
A simple explanation for a beautiful experiment}

\vskip 2.2cm
\centerline{R. Foot\footnote{
E-mail address: rfoot@unimelb.edu.au}}

\vskip 0.7cm
\centerline{\it School of Physics,}
\centerline{\it University of Melbourne,}
\centerline{\it Victoria 3010 Australia}
\vskip 2cm
\noindent
Recently, the  DAMA/LIBRA experiment has convincingly confirmed  
the DAMA/NaI annual modulation signal, 
experimentally demonstrating the existence of non-baryonic dark matter in the halo of our galaxy. 
Meanwhile, in  another part  of town, 
other experiments such as CDMS and XENON10 have  not
detected  any evidence for dark  matter. 
One promising dark matter candidate 
which can reconcile the positive DAMA annual modulation
signal with the  null  results from  the other  experiments, is  mirror dark matter.
We re-analyse the mirror matter interpretation of the  DAMA annual modulation signal 
utilizing a) the
new data from DAMA/LIBRA, including the measured energy dependence of the
annual modulation signal b) an updated quenching factor which takes into account
the channeling  effect  in  $NaI$ crystals and c) the latest constraints
from CDMS/Ge, CDMS/Si and XENON10 experiments.
We show that the simplest possibility of a  $He'$ (and/or $H'$) dominated  halo with
a small $O'$ component is sufficient to fully explain all  of the  dark 
matter experiments. We also point out that a certain class of 
hidden sector dark matter models, although theoretically less
appealing and less constrained, can mimic the success of the mirror dark matter model
and hence are also viable.

\end{titlepage}

\section{Introduction}

Recently the DAMA/LIBRA experiment has confirmed\cite{damalibra} the impressive DAMA/NaI annual
modulation signal at around $8\sigma$ C.L.\cite{dama}. Their signal is observed in the 2-6 keVee energy region, 
with periodicity  and phase:
\begin{eqnarray}
T/{\rm year} = 0.998 \pm 0.003, \ t_0/{\rm day} =  144 \pm 8,
\end{eqnarray}
which  is beautifully consistent with the  dark matter expectation
of  $T/{\rm year} =  1.0$,  $t_0/{\rm day} = 152.5$. 
There are no known systematic effects
which could produce the modulation  of the  signal seen. 
Indeed, 
if there were some hypothetical systematic effect producing the modulation,
it would be surprising
if it had the  same periodicity AND  phase  as the
dark  matter  signal. 
Therefore, it is reasonable to believe that these experiments have detected galactic dark matter particles.


Meanwhile other experiments have failed to detect positive evidence for dark matter.
Among the most sensitive of these, are the recent 
CDMS/Ge\cite{cdms2}, CDMS/Si\cite{cdms1} and XENON10\cite{xenon} experiments. 
Combined analysis of these null results with the positive results of DAMA appear
to  exclude  standard  WIMP  explanations  (such  as  the standard neutralino  models) for
the DAMA  signal. 
Several alternative explanations of  the  DAMA signal have  been proposed, such as the elastic 
scattering of  light WIMPs (with mass $\sim$ 7 GeV)\cite{lightwimps}, 
and very light candidates (e.g. with
mass in the keV range) scattering  off  electrons\cite{lightdm1,lightdm2}.
While such phenomological  models
might be possible, it is  a theoretical challenge to construct simple renormalizable
extensions to the standard model which can accommodate these light particles.

On the other hand, previous work\cite{f1,f2,f3,f4} has shown that mirror
dark matter can explain the DAMA/NaI annual modulation signal, consistently
with the null results of the other experiments.
Mirror dark matter arises in simple renormalizable extensions of the standard model
featuring a sector of particles and forces {\it exactly}
isomorphic (or ``mirror'') to the known particles and forces. 
The simplest such model being the exact parity symmetric model\cite{flv} 
(for a review, see e.g. ref.\cite{review}). 
 These theories feature a spectrum of dark matter particles, $A'$,
with known masses. 

The success of the mirror dark matter theory in explaining 
the DAMA/NaI experiment consistently
with the null results of the other dark matter detection experiments has to do with several
of its key distinctive features:
\begin{itemize}
\item Galactic halo mirror particles have 
a Maxwellian velocity distribution: $f(u) = 
exp[-u^2/v_0^2]$,
however the velocity dispersion of heavy mirror particles ($M_{A'} > M_{He}$) 
is much less than the galactic rotational velocity: $v_0(A')^2 \ll v_{rot}^2$. In standard
WIMP models, $v_0^2 = v_{rot}^2$. 
\item The mirror dark matter mass spectrum consists of relatively light
particles: $M_{A'} \le M_{Fe}$. 
Standard WIMP models typically feature $M_{WIMP} \stackrel{>}{\sim} 50$ GeV.
\item Mirror dark matter interacts with ordinary 
matter via renormalizable photon-mirror photon kinetic mixing, leading to a
recoil energy ($E_R$) dependent cross-section: 
${d\sigma \over dE_R}\propto {1 \over E_R^2}$.  
Standard WIMP models have an approximately $E_R$ independent cross-section arising
from a four-Fermion interaction.
\end{itemize}
These three features greatly increase the sensitivity of mirror dark matter
experiments such as DAMA/NaI. The main reason is that the DAMA/NaI experiment
is a lower threshold experiment than the sensitive CDMS and Xenon experiments,
when the channeling effect is taken into account\cite{damaquench}.

The  purpose  of  this  paper is to re-analyse the  mirror
matter interpretation  of  the DAMA  annual modulation signal in the light of the new
data from  DAMA/LIBRA  as  well  as the latest constraints  from  CDMS and XENON  
experiments. We also use  an  updated quenching factor, taking  into  account
the channeling effect in  $NaI$  crystals, which was discussed recently  by
the DAMA collaboration\cite{damaquench}.

The outline of this paper is as follows: in section II we give a brief review
of mirror dark matter, which may be skipped by those already familiar with the subject.
In section III, we give the necessary details 
relevant to direct detection experiments such as cross-section, halo distribution function etc.
In section IV we give the detailed fit to the DAMA/LIBRA + DAMA/NaI combined results under 
the simple and plausible assumption that the heavy component of the galactic halo (i.e. with $M > M_{He}$) 
is dominated by just one mirror element. 
That is, one element in addition to the $H'/He'$ predominant component.
In section V we examine the constraints
imposed by the negative results of the CDMS/Ge, CDMS/Si and XENON10 experiments.
In section VI we point out that a class of hidden sector models can mimic 
the success of the mirror dark matter theory in explaining the experiments.
Finally, in section VII we draw our conclusions.
 
\section{Mirror Dark Matter}

The exact parity symmetric model\cite{flv} is the minimal extension of
the standard model which allows for an exact unbroken parity symmetry
[$x \to -x$, $t \to t$].
According to this theory, each type of ordinary
particle (electron, quark, photon etc) has a corresponding mirror partner
(mirror electron, mirror quark, mirror photon etc), 
of the same mass. The two sets of particles form 
parallel sectors each with gauge symmetry $G$
(where $G = SU(3) \otimes SU(2) \otimes U(1)$ in the 
simplest case)
so that the full gauge group is $G \otimes G$.
The unbroken mirror symmetry maps
$x \to -x$ as well as ordinary particles into mirror
particles. Exact unbroken time reversal symmetry
also exists, with standard CPT identified as the product
of exact T and exact P\cite{flv}.

It has been argued that the stable mirror particles, mirror nucleons and
mirror electrons are an interesting candidate for the inferred
dark matter of the Universe (for a review, see Ref.\cite{review}).
Of course, to be a successful dark matter candidate, mirror
matter needs to behave, macroscopically, differently to ordinary matter.
In particular, four key distinctions need to be explained:
\begin{itemize}
\item
The cosmological abundance of mirror matter should be different to
ordinary matter,
$\Omega_{dark} \neq \Omega_{matter}$. 
\item
Mirror particles
should give negligible contribution to the energy density at the
epoch of big bang nucleosynthesis.
\item
Structure formation in the mirror
sector must begin before ordinary matter radiation decoupling.
\item
In spiral galaxies,
the time scale for the collapse of mirror matter onto the disk
must be much longer than that of ordinary matter.

\end{itemize}

Clearly, mirror matter behaves differently to ordinary matter,
at least macroscopically.
It is hypothesised that
this macroscopic  asymmetry originates from
effectively
different initial conditions in the two sectors. The exact
microscopic (Lagrangian) symmetry between ordinary and mirror
matter need never be broken. In particular, if ordinary
and mirror particles have different temperatures in the
early Universe, $T' \ll T$,\footnote{
It has been speculated that $T' \ll T$ might arise in certain
inflation scenarios, see e.g. ref.\cite{kst}.} then
the mirror particles give negligible contribution to the energy
density at the time of nucleosynthesis leading to standard big
bang nucleosynthesis. Another consequence of $T' \ll T$ is that
mirror photon decoupling occurs earlier than ordinary photon 
decoupling -- implying that mirror structure formation can begin
before ordinary photon decoupling. 
In this way, mirror matter-type dark matter can successfully explain
the large scale structure formation (for detailed studies, 
see ref.\cite{comelli,lss,other}).  
Also, $\Omega_{dark} \neq
\Omega_{matter}$ could also be due to different effective initial conditions
in the early Universe (see e.g. ref.\cite{new1},\cite{new2} for some
specific scenarios).

If mirror matter is the inferred non-baryonic dark matter
in the Universe, then
the halo of our galaxy
should be a gas of ionized mirror atoms and mirror electrons with a
possible mirror star component\footnote{The mirror star component 
can be probed by microlensing observations towards nearby galaxies\cite{sil}.
The current situation is somewhat unclear, but a mass fraction of MACHO's
around $f \sim 0.1$ is roughly compatible with the observations. For a
review, see ref.\cite{reviewmacho}}.
Although dissipative, roughly spherical galactic mirror matter halo's can exist without
collapsing provided that a heating mechanism exists - with
ordinary supernova explosions being plausible
candidates\cite{sph}. 
Obviously, the heating of the ordinary and mirror matter
in spiral galaxies needs to be asymmetric, but again, due to
different initial conditions in the early Universe, asymmetric heating
is plausible. For example, the early Universe temperature asymmetry, 
$T' \ll T$ (expected from successful big bang nucleosynthesis and Large
scale structure formation, as discussed above) 
implies that the primordial mirror
helium/mirror hydrogen ratio will be much larger than the corresponding
ordinary helium/ordinary hydrogen ratio\cite{comelli}. 
Consequently the formation and evolution of stars in the ordinary
and mirror sectors are completely different (see e.g. ref.\cite{diff} for 
some preliminary studies).
The details of
the evolution on (sub) galactic scales, is of course, very complex, and is  
yet to be fully understood. 

Ordinary and mirror particles interact with each other
by gravity and via renormalizable photon-mirror photon kinetic
mixing:\footnote{
Technically, photon-mirror photon kinetic mixing arises from
kinetic mixing of the $U(1)$ and $U(1)'$ gauge fields, since
only for the abelian $U(1)$ gauge symmetry is such mixing
gauge invariant\cite{fh}.
The only other gauge invariant and renormalizable interactions mixing
ordinary and mirror particles are the quartic Higgs - mirror Higgs interaction:
$\lambda \phi^{\dagger}\phi \phi'^{\dagger}\phi'$ and neutrino - mirror
neutrino mass mixing\cite{flv,flv2}.}
\begin{eqnarray}
{\cal L} = {\epsilon \over 2}
F^{\mu \nu} F'_{\mu \nu}
\end{eqnarray}
where $F^{\mu \nu}$ ($F'_{\mu \nu}$) is the field
strength tensor for electromagnetism (mirror electromagnetism).
One effect of photon-mirror photon kinetic mixing is to cause mirror
charged particles (such as the mirror proton and mirror electron)
to couple to ordinary photons with effective electric charge $\epsilon e$.\cite{flv,holdom,sasha}
The various experimental implications of photon-mirror photon kinetic 
mixing have been
reviewed in Ref.\cite{freview}.
Of most relevance for this paper, is that
this interaction enables mirror particles 
to elastically scatter off ordinary particles -- essentially
Rutherford scattering.

A detector on Earth can therefore be used to detect halo mirror
nuclei via elastic scattering. Several previous
papers\cite{f1,f2,f3,f4} have explored this possibility, especially in view
of the impressive dark matter signal from the DAMA/NaI experiment. 
The purpose of this paper is to re-analyse the mirror matter interpretation of
the DAMA/NaI experiment using the new  data  from DAMA/LIBRA and  also
the latest  constraints from the null results of the CDMS and XENON10 experiments.
In  the following section we  give the technical details - the cross section and
halo distribution  function - necessary to facilitate this comparison 
of theory  and  experiment.

\section{Mirror dark matter implications for direct detection 
experiments}

\subsection{Cross-section and form factors}

Let us first briefly review the required technology (see ref.\cite{f1,f2,f3,f4} for
more details).
For definiteness, consider
a halo mirror nuclei, $A'$, of atomic number $Z'$
scattering off an ordinary nucleus, $A$ (in an ordinary matter detector) 
of atomic number $Z$.
The cross-section is then just of the standard Rutherford form
corresponding to a particle of electric charge $Ze$ scattering
off a particle of electric charge $\epsilon Z' e$.
This cross-section can be expressed in terms of the recoil
energy of the ordinary atom, $E_R$, and 
the velocity of $A'$ in the Earth's rest frame, $v$:
\begin{eqnarray}
{d\sigma \over dE_R} = {\lambda \over E_R^2 v^2}
\label{cs}
\end{eqnarray}
where 
\begin{eqnarray}
\lambda \equiv {2\pi \epsilon^2 \alpha^2 Z^2 Z'^2 \over M_A} \
F_{A}^2 (qr_A) F_{A'}^2 (qr_{A'})
\end{eqnarray}
and $F_X (qr_X)$ ($X = A, A'$) are the form factors which 
take into account the finite size of the nuclei and mirror nuclei.
[$q = (2M_A E_R)^{1/2}$ is the momentum transfer and $r_X$ is the effective
nuclear radius]\footnote{
We use natural units, $\hbar = c = 1$ throughout.}.
A simple analytic expression for
the form factor, which we adopt in our numerical work, is the one
given by Helm\cite{helm,smith}:
\begin{eqnarray}
F_X (qr_X) = 3{j_1 (qr_X) \over qr_X} e^{-(qs)^2/2}
\end{eqnarray}
with $r_X = 1.14 X^{1/3}$ fm, $s = 0.9$ fm.
In this equation, $j_1$ is the spherical Bessel function of index 1.

\subsection{Interaction rate and halo distribution function}

In an experiment such as DAMA/NaI, the measured quantity is 
the recoil energy, $E_R$, of a target atom. 
The interaction rate is
\begin{eqnarray}
{dR \over dE_R} &=& 
\sum_{A'} N_T n_{A'} \int {d\sigma \over dE_R}
{f_{A'}(\textbf{v},\textbf{v}_E) \over k} |v|
d^3v \nonumber \\
&=& \sum_{A'} N_T n_{A'}
{\lambda \over E_R^2 } \int^{\infty}_{|v| > v_{min}
(E_R)} {f_{A'}(\textbf{v},\textbf{v}_E) \over k|v|} d^3 v 
\label{55}
\end{eqnarray}
where $N_T$
is the number of target atoms per kg of detector\footnote{
For detectors with more than one target element 
we must work out the
event rate for each element separately and add them up to get the total
event rate.}. Also,
$n_{A'}$ is the halo number density (at the Earth's location) 
of the mirror element, $A'$ and 
$f_{A'}(\textbf{v},\textbf{v}_E)/k$ is its velocity distribution 
($k \equiv (\pi v_0^2)^{3/2}$ is the normalization factor).
Here, $\textbf{v}$ is the velocity of the halo particles relative
to the Earth, and $\textbf{v}_E$ is the Earth's velocity relative to the
galactic halo.
[Note that the bold font is used to indicate that the quantities are
vectors].
The lower velocity limit,
$v_{min} (E_R)$, 
is given by the kinematic relation:
\begin{eqnarray}
v_{min} &=& \sqrt{ {(M_A + M_{A'})^2 E_R\over 2M_A M^2_{A'}} } .
\label{v}
\end{eqnarray}

Considering a particular mirror chemical element, $A'$ (e.g.
$A' = H', He', O'$ etc), the velocity distribution for
these halo mirror particles is then:
\begin{eqnarray}
f_{A'} (v, v_E) = exp\left[ -{1 \over 2} M_{A'} (\textbf{v}+\textbf{v}_E)^2/T \right]
= exp[-(\textbf{v}+\textbf{v}_E)^2/v_0^2]
\label{d12}
\end{eqnarray}
where $v_0^2 \equiv 2T/M_{A'}$. 

The assumption of approximate
hydrostatic equilibrium for the halo particles implies
a relation between $T$ and the local rotational velocity, 
$v_{rot}$:\cite{f2}
\begin{eqnarray}
T = {\mu M_p v_{rot}^2 \over 2}
\label{d13}
\end{eqnarray}
where $\mu M_p$ is the mean mass of the particles
comprising the mirror (gas) component of the halo ($M_p$ is the
proton mass).
Note that the Maxwellian distribution should be an excellent
approximation in the case of mirror dark matter, since
the self interactions of the particles ensure that the halo is
locally thermalized.

The velocity integral in Eq.(\ref{55}),
\begin{eqnarray}
I (E_R) \equiv \int^{\infty}_{|v| > v_{min}(E_R)}
{f_{A'}(\textbf{v},\textbf{v}_E) \over k|v|} d^3 v
\label{sat}
\end{eqnarray}
is standard (similar integrals occur in the usual
WIMP interpretation\footnote{
However in the WIMP case the upper velocity limit is finite, 
corresponding to the galactic escape velocity. While for
mirror dark matter, the upper limit is infinite due
to the self interactions of the mirror particles.}) and can 
easily be evaluated in terms of
error functions assuming
a Maxwellian dark matter distribution\cite{smith},
$f_{A'}(\textbf{v},\textbf{v}_E)/k = (\pi v_0^2)^{-3/2} \ exp[-(\textbf{v}+\textbf{v}_E)^2/v_0^2]$,
\begin{eqnarray}
I(E_R) = {1 \over 2v_0 y}\left[ erf(x+y) - erf(x-y)\right] 
\label{ier}
\end{eqnarray}
where 
\begin{eqnarray}
x \equiv {v_{min} (E_R) \over v_0}, \ y \equiv {v_E \over v_0}.
\label{xy}
\end{eqnarray}
The Earth's velocity relative to the galactic halo, $v_E$, has
an estimated mean value  of
$\langle v_E \rangle \simeq v_{rot} + 12 \ {\rm km/s}$, 
with  $v_{rot}$, the local
rotational velocity, in the
$90\%$ C.L. range\cite{koch},
\begin{eqnarray}
170 \ {\rm km/s} \stackrel{<}{\sim} v_{rot} \stackrel{<}{\sim} 270\ {\rm km/s}.
\label{range2}
\end{eqnarray}
While some estimates put narrower limits on the local
rotational velocity, it is useful to allow for
a broad range for $v_{rot}$ since it can also
approximate the effect of bulk halo rotation.

As can be seen from Eq.(\ref{d12},\ref{d13}),
in the case of mirror matter-type dark matter,
the $v_0$ value
for a particular halo component element, $A'$,
depends on the chemical composition of the halo.
In general,
\begin{eqnarray}
{v_0^2 (A') \over v_{rot}^2} = {\mu M_p \over M_{A'}}\ .
\label{z3}
\end{eqnarray}
The most abundant mirror elements
are expected to be $H', He'$, generated in
the early Universe from mirror big bang nucleosynthesis (BBN).
Heavier mirror elements are expected to be generated in mirror
stars, or possibly, in the early universe if mirror BBN occurs
early enough, so  that the number densities are large enough, 
to allow three-body 
processes such as the triple ${\rm alpha}'$ process
to be effective. It is useful, therefore, to consider
two limiting cases: first that the halo
is dominated by $He'$ and the second is that
the halo is dominated by $H'$. The mean mass of
the particles in the halo are then
(taking into
account that the light halo mirror atoms should be fully ionized):
\begin{eqnarray}
\mu M_p & \simeq  & M_{H'}/2 \simeq 0.5\ {\rm GeV \ for \ H' \ dominated \ halo,}\nonumber \\
\mu M_p & \simeq & M_{He'}/3 \simeq 1.3\ {\rm GeV \ for \ He' \ dominated \ halo.}
\end{eqnarray}
The $v_0$ values can then easily be obtained from Eq.(\ref{z3}):
\begin{eqnarray}
v_0 (A') &=& v_0 (H') \sqrt{{M_{H'}\over M_{A'}}} \approx {v_{rot}
\over \sqrt{2}}
\sqrt{{M_{H'}\over M_{A'}}} \ {\rm km/s} \ \ \ \rm{for \ H' \ dominated \
halo}
\nonumber \\
v_0 (A') &=& v_0 (He') \sqrt{{M_{He'}\over M_{A'}}} \approx {v_{rot} \over
\sqrt{3}}
\sqrt{{M_{He'}\over M_{A'}}} \ {\rm km/s} \ \  \ \rm{for \ He' \ dominated \
halo}.
\end{eqnarray}
Mirror BBN\cite{comelli}
suggests that $He'$ dominates over $H'$. 
It happens 
that the recoil energy thresholds of the DAMA/NaI and the other experiments
are sufficiently high that these experiments are only directly sensitive to
mirror elements heavier than about carbon, which 
means that
$v_0 (A') \ll v_{rot}$ for these elements -- independently of 
whether $He'$ or $H'$ dominates the halo.
It  turns out that our main results 
(such as the allowed region in figure 4)
do not depend very significantly
on whether we assume that $He'$ or $H'$ dominates the mass
of the Halo.

Finally, we see from Eq.(\ref{55}) that  the interaction rate,  
$dR/dE_R$, is  proportional to $\epsilon^2 n_{A'}$, where
$\epsilon$ is the photon-mirror photon kinetic mixing parameter and $n_{A'}$ is the halo number 
density of the species  $A'$.  The  dark matter halo  mass density at the Earth's 
location is approximately $0.3\ GeV/cm^3$,
so that the halo  mass  fraction, $\xi_{A'}$, of the species  
$A'$, is thus: $\xi_{A'} \equiv n_{A'} M_{A'}/(0.3\ GeV/cm^3)$.
Hence, we find that the interaction rate is proportional  to $(\epsilon \sqrt{\xi_{A'}})^2$. 
We call the quantity, $\epsilon \sqrt{\xi_{A'}}$, the 
cross-section abundance coefficient, which we  can determine from
the amplitude of the DAMA  annual  modulation signal. The only other 
variables  in  the model  are the  rotation velocity,  $v_{rot}$,  which
is  expected
to be in  the  range,  Eq.(\ref{range2}),  and  the spectrum of heavy mirror elements,
$A'$.  

\section{Mirror matter interpretation 
of DAMA/NaI \& DAMA/LIBRA annual modulation
signal}

The DAMA experiments employ large mass scintillation detectors located in the Gran Sasso
underground laboratory. The first generation, DAMA/NaI experiment ran for 7 years,
presenting final results in 2003\cite{dama}. The second generation, DAMA/LIBRA experiment
has been running since 2003, recently presenting its first results from
4 years of data\cite{damalibra}.

The DAMA/NaI and DAMA/LIBRA experiments are very sensitive to mirror
matter-type dark matter because 
of the relatively low energy threshold of 2 keVee.
The unit, keVee is the so-called electron equivalent energy,
which is the energy of the event if due to an electron
recoil. The actual nuclear recoil energy
(in keV) is given by: keVee/$q$, where $q$ is the quenching factor.

The DAMA/NaI and DAMA/LIBRA experiments extract their signal by using 
the annual modulation signature\cite{sig},
which arises because of the Earth's orbital motion. 
The point is that the interaction rate, Eq.(\ref{55}), depends on $v_E$,
which varies due to the Earth's motion around the sun:
\begin{eqnarray}
v_E (t) &=& v_{\odot} + v_{\oplus} \cos\gamma \cos \omega (t - t_0)
\nonumber \\
&=& v_{\odot} + \Delta v_E \cos \omega (t - t_0)
\end{eqnarray}
where $v_{\odot} = v_{rot} + 12 \ {\rm km/s} \sim 230 \ {\rm km/s}$ is the sun's velocity
with respect to the galactic halo and $v_{\oplus} \simeq 30$ km/s is the Earth's orbital
velocity around the Sun ($t_0 = 152.5$ days and 
$\omega = 2\pi/T$, with $T = 1$ year).
The inclination of the Earth's orbital plane relative
to the galactic plane is $\gamma \simeq 60^o$, which implies that 
$\Delta v_E \simeq 15$ km/s.
The differential interaction rate, Eq.(\ref{55}), 
can be expanded in a Taylor series around $v_E = v_{\odot}$, leading
to an annual modulation term:
\begin{eqnarray}
R_i = R^0_i + R^1_i \cos \omega (t - t_0)
\end{eqnarray}
where
\begin{eqnarray}
R^0_i &=& {1 \over \Delta E} \int^{E_i + \Delta E}_{E_i} \left( {dR \over dE_R}\right)_{v_E = v_{\odot}} \ dE_R,
\nonumber \\
R^1_i &=& {1 \over \Delta E} \int^{E_i + \Delta E}_{E_i} {\partial \over \partial v_E} \left(
{dR \over dE_R}\right)_{v_E = v_{\odot}} \ \Delta v_E dE_R
\label{fri}
\end{eqnarray}

According to the DAMA analysis\cite{damalibra}, they indeed find an annual 
modulation at more than $8\sigma$ C.L.
Their data fit gives $T = (0.998 \pm 0.003)$ years and $t_0 = 144 \pm 8$ days, 
consistent
with the expected values. [The expected value for $t_0$ is 152.5 days (2 June), where
the Earth's velocity reaches a maximum with respect to the galactic halo].
Their signal occurs in the 2-6 keVee energy range, with amplitude
\begin{eqnarray}
R^1 = (0.0129 \pm 0.0016) \ {\rm cpd/kg/keVee} \  {\rm [cpd = counts \ per \ day]}.
\label{kj}
\end{eqnarray}

The theoretical  rate for the modulated and
absolute component can be evaluated  in  our  model from
Eqs.(\ref{fri},\ref{55},\ref{sat},\ref{ier}),
leading to:
\begin{eqnarray}
{dR^0 \over dE_R} &=& \sum_{A'} {N_T n_{A'} \lambda  I(E_R,y_0) \over  E^2_R} \nonumber \\
{dR^1 \over dE_R} &=& \sum_{A'} {N_T n_{A'} \lambda \Delta y \over E_R^2} \left(
{\partial I \over \partial y}\right)_{y=y_0}
\end{eqnarray}
where $y_0 = \langle v_E  \rangle/v_0$,  $\Delta y = \Delta  v_E/v_0$, and
\begin{eqnarray}
\left({\partial I \over \partial y}\right)_{y=y_0} =  - {I(E_R,y_0) \over y_0} + 
{1 \over \sqrt{\pi} v_0 y_0} \left[ e^{-(x-y_0)^2} + e^{-(x+y)^2}\right] \ .
\end{eqnarray}
Recall, from Eqs.(\ref{xy},\ref{v},\ref{z3}) that 
$x \propto \sqrt{E_R}$, while $y_0 \simeq {v_{rot} \over v_0} =
\sqrt{{M_{A'} \over \mu M_p}}$,
so that $y_0 \simeq 3.4\sqrt{{M_{A'} \over M_{O'}}}$ $\left(5.5 \sqrt{{M_{A'} \over
M_{O'}}}\right)$ 
for a $He'$ ($H'$) dominated halo.

Consider the interactions of  one halo component $A'$.  What is  
the behaviour  of  the annual modulation amplitude, 
$dR^1/dE_R$, as a function of  $E_R$?
At low  $E_R$, (where $x  \ll y$),  $dR^1/dE_R$ is  negative, as  $E_R$ increases,
$dR^1/dE_R$ changes sign, and reaches a maximum when 
\begin{eqnarray}
x(E_R) \simeq y_0.
\label{peak}
\end{eqnarray} 
At  high $E_R$ ($x \gg y$), $dR^1/dE_R \to 0$.

Of course,  in  order to do a detailed fit of the experimental data to a given theoretical model,
we must take into account the finite resolution of the DAMA detector and the 
physics of the quenching factor.
In all of our numerical work both of these effects are included.
The experimental energy resolution is taken into account by 
convolving the rate with
a Gaussian, with $\sigma$ obtained 
from ref.\cite{recent2}: $\sigma/E = \alpha/\sqrt{E (keV)} + \beta$
with $\alpha = 0.448 \pm 0.035, \ \beta = (9.1 \pm 5.1)\times 10^{-3}$.

The quenching factor is more complicated. Our previous work
assumed $q_{Na} = 0.3$ and $q_{I} = 0.09$. However, recently 
Drobyshevski\cite{dob} has pointed out the possible importance of 
the `channeling effect' in $NaI$ crystals,
which means that ions moving approximately in the direction of crystal axes and planes
get `channeled' whereby they give an unquenched signal, $q\simeq 1$.
A detailed study and modelling of this effect has been done recently
by the DAMA collaboration\cite{damaquench}.
According to their model, the fraction 
of events, $f_{Na,I}$, with quenching factor $q \simeq 1$, 
in the low recoil energy region of interest is approximately: 
\begin{eqnarray}
f_{Na} \simeq {1 \over 1 + 1.14 E_R (keV)}, \ \ 
f_I \simeq {1 \over 1 + 0.75 E_R (keV)}
\label{yay}
\end{eqnarray}
The remaining fraction of events have quenching factors, $q_{Na} \sim 0.3,\ q_I \sim 0.09$ (but
are broadened by straggling).
The channeling effect is very important, as it effectively reduces the threshold of the DAMA
experiments down to an actual recoil energy of 2 keV. It turns out that the fitted cross-section is
over an order of magnitude smaller compared to the case where the channeling effect is ignored.
This means that 
the interaction rate is dominated by the channeled  events.

The 2 keV threshold is still too high for the DAMA experiment to be  very sensitive to the dominant, 
$He'$ or $H'$ component, except possibly for effects in the $E_R \stackrel{<}{\sim} 2.5$ keV
region from the tail of the $He'$ induced events.
Basically, these nuclei are too light to give much of a signal 
above the DAMA/NaI energy threshold. [Of course,
the DAMA experiment is indirectly sensitive to the $H'/He'$ component, which as 
we discussed in section 3, has
the important effect of reducing the value of $v_0 (A')$ ].
DAMA is only directly sensitive to mirror nuclei heavier than about carbon.
In previous work we considered having several components, such as $O', Si', Fe'$ in ref.\cite{f4},
however it turns out that this is not necessary to fit the new DAMA/LIBRA data.
The reason is two fold: first, DAMA/LIBRA have found a lower annual modulation amplitude
in the 4-6 keVee energy region
than was found in the earlier DAMA/NaI experiment, second,
the channeling effect also modifies the predicted spectrum. 
The net effect, as we will see, is that the interaction of a 
single halo element predicts a signal
of the correct shape in the 2-6 keVee range.
We therefore assume this simplest possibility, that is, that the
spectrum of heavy elements (in this context `heavy' means heavier than $M_{He}$) is dominated by just
one element, $A'$. This would also approximate a narrow range
of elements.  The detailed predictions will depend on the mass of the
element $A'$, the rotational velocity $v_{rot}$ and the cross-section abundance coefficient,
$\epsilon \sqrt{\xi_{A'}}$. 

Note that the cross-section abundance coefficient,
$(\epsilon \sqrt{\xi_{A'}})^2$, is just an overall normalization factor, and doesn't change
the shape of the annual modulation energy spectrum. This factor can be adjusted so that
$R^1$ fits the measured  amplitude averaged  over the 2-6 keVee energy range. 
Of course, only the
amplitude is a free parameter, the phase, periodicity and the approximate cosine form of the signal  are
all predicted, as they are for many other dark  matter models.
The fit of the annual modulation data is shown in {\bf figure 1}.

\vskip 0.6cm

\centerline{\epsfig{file=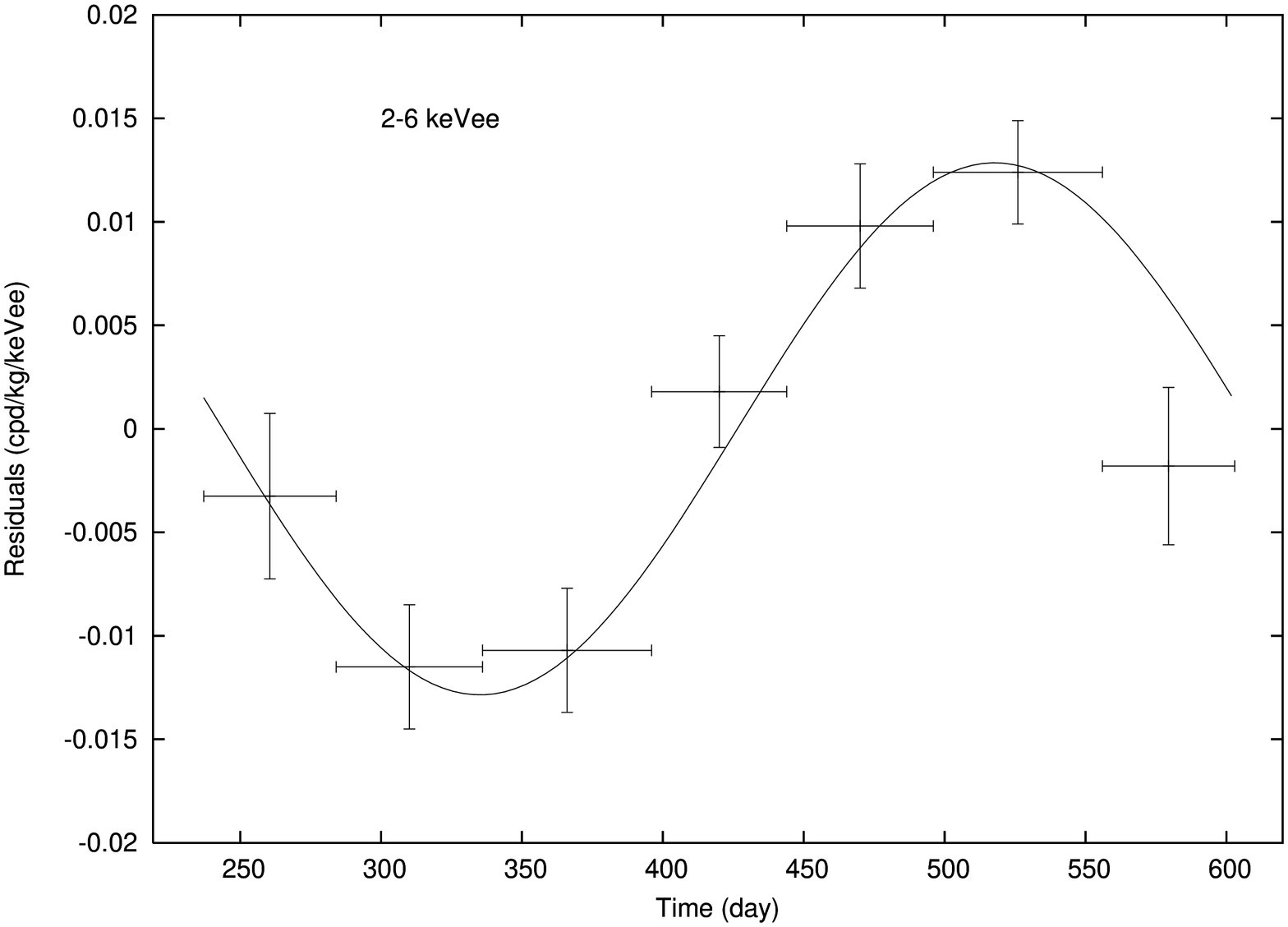,angle=270,width=12cm}}

\vskip 0.5cm

\noindent
Figure 1: Measured single hit residuals for the (2-6) keVee region
(obtained from figure 6 or ref.\cite{damalibra})
compared to the predicted cosine modulation with 
phase $t_0/{\rm day} = 152.5$ day
and $T/{\rm year}$ = 1. 

\vskip 1cm

An important new result from DAMA/LIBRA is that we now have a much better idea of the shape
of the annual modulation energy spectrum.
The DAMA collaboration separate the data in the 2-6 keVee range into 8 energy bins
and fit the modulation in each energy bin to the predicted cosine distribution, obtaining
modulation amplitudes, $S_i^m \pm \sigma_i$, for each energy bin. These amplitudes  can be
compared to the theoretically  computed quantities, $R^1_i$,
taking into account the
experimental resolution and quenching factors as described earlier. We can
compute a $\chi^2$ quantity, 
\begin{eqnarray}
\chi^2 = \sum_{i=1}^8 (S_i^m - R^1_i)^2/\sigma_i^2
\label{chi2}
\end{eqnarray}
which  is a function of  
the three parameters: $v_{rot}, \ \epsilon \sqrt{\xi_{A'}}$ and the identity of the dominant
element $A'$. Thus, with eight independent data bins we have  five degrees of freedom.
We find numerically that for
each value of $v_{rot}$ [in the  range, Eq.(\ref{range2})], the model 
provides a  good fit to the data, with:
\begin{eqnarray}
\chi^2_{min} &\approx & 5.5, \ \ {\rm for \ H' \ dominated \ halo} \nonumber \\ 
\chi^2_{min} &\approx & 3.5, \ \ {\rm for \ He' \ dominated \ halo} 
\end{eqnarray}
for five degrees of freedom.
The data does constrain the other two parameters.
Numerically we obtain the $3\sigma$ range for  $\epsilon \sqrt{\xi_{A'}}$:
\begin{eqnarray}
|\epsilon | \sqrt{ {\xi_{A'} \over 0.1}}
&\simeq & (1.0 \pm 0.3) \times 10^{-9} 
\left({v_{rot} \over 210\ {\rm km/s}}\right)^{{7 \over
4}}
\label{dama55}
\end{eqnarray}
for both $H'$ and $He'$ dominated halo.

The  allowed  range  for $A'$ also  depends on $v_{rot}$. We find that we can
fit the  DAMA data for $A'$ ranging in mass from oxygen to sulfur, i.e. 
$15$ GeV $\stackrel{<}{\sim} M_{A'} \stackrel{<}{\sim} 30$ GeV.
In  {\bf figure 2} we have given 
the best fits for four illustrative examples, $A' = 0', \ v_{rot} = 280$ km/s,
$A' = Ne', \ v_{rot} = 230$ km/s, $A' = Mg', \ v_{rot} = 195$ km/s
and $A' = Si', \ v_{rot} = 170$ km/s. 
Figure 2a assumes a $H'$ mass dominated halo, while figure 2b assumes a $He'$ mass dominated halo.

\vskip 1cm

\centerline{\epsfig{file=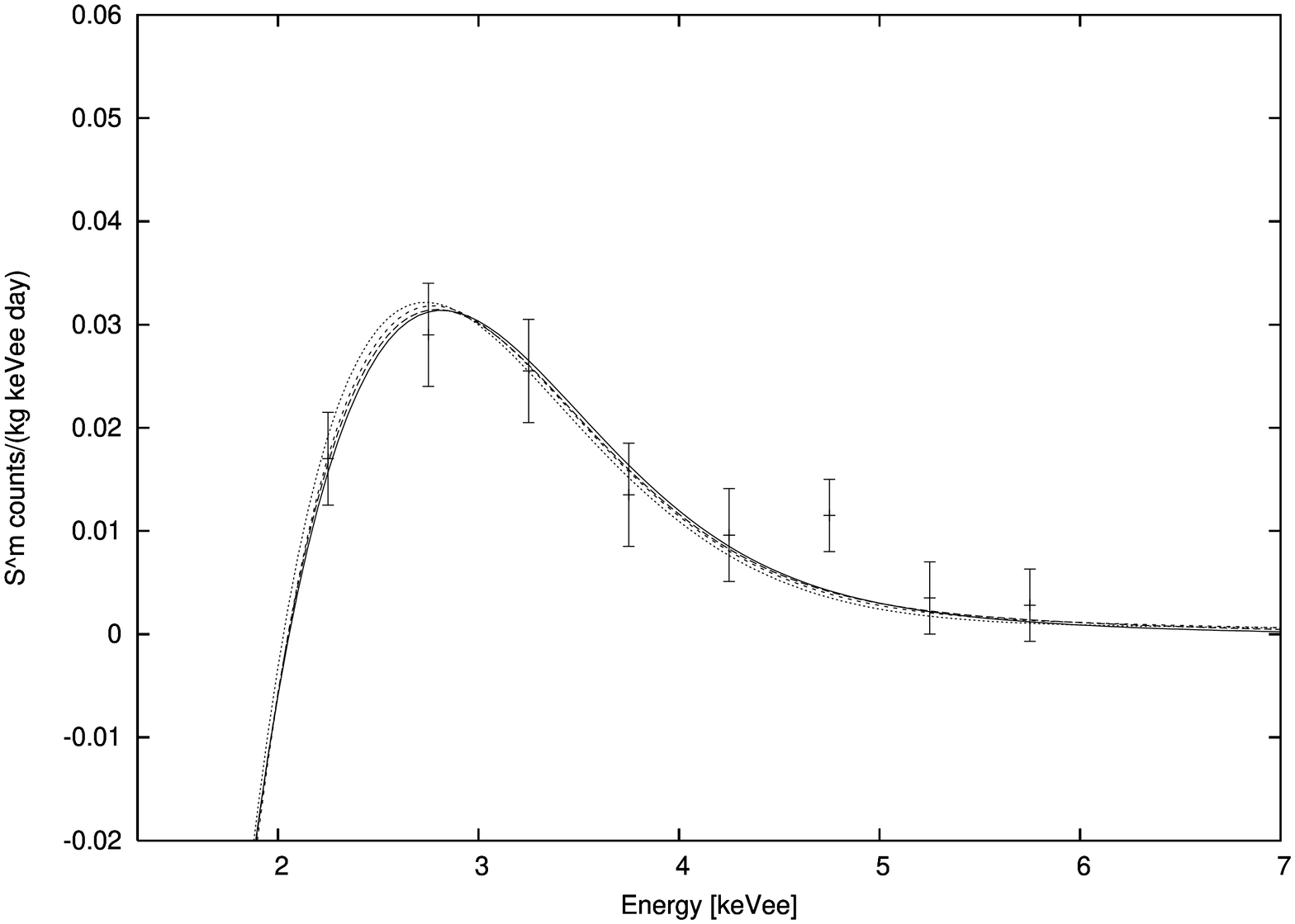,angle=270,width=13cm}}
\vskip 0.5cm
\noindent
Figure 2a:
Energy distribution of the cosine modulation amplitude for four illustrative cases: 
$A'=Si', v_{rot}=170$ km/s (solid line)
$A'=Mg', v_{rot}=195$ km/s (long-dashed line),
$A'=Ne', v_{rot}=230$ km/s (short-dashed line),
$A'=O', v_{rot}=280$ km/s (dotted line).
In each case $\epsilon \sqrt{\xi_{A'}}$ is fixed so that the mean amplitude is 0.0129 cpd/kg/keVee.
Also shown are the DAMA/NaI \& DAMA/LIBRA combined data from figure 9 of ref.\cite{damalibra}.
This figure assumes that the mass of the halo is dominated by $H'$.

\vskip 1cm
\centerline{\epsfig{file=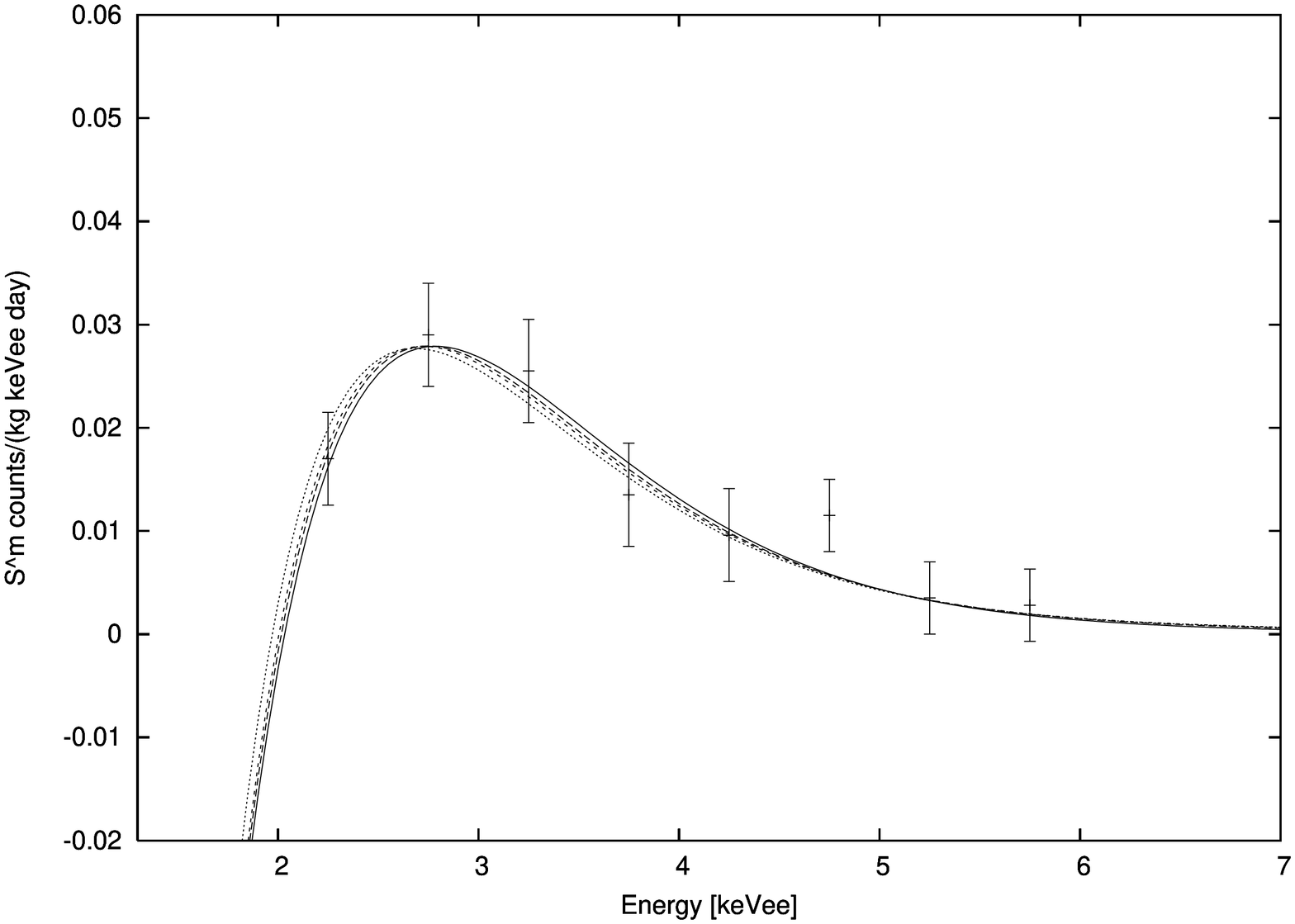,angle=270,width=13cm}}
\vskip 0.5cm
\noindent
Figure 2b: Same as figure 2a, except the mass of the halo is assumed to be dominated by $He'$.

\vskip 1cm

The shape of the measured annual modulation energy spectrum agrees very 
well with the  mirror  dark matter theory.
It is  essentially  a  two parameter  fit, since  as figure 2 demonstrates, there  is  an
approximate redundancy.  Once the  height  and  position  of the peak
is  fitted, there is  little  freedom  left to  change  the  shape  of the  distribution.
Thus, the fact that the measured shape agrees with that predicted by 
the mirror matter theory is a significant test of the theory.
We illustrate this in {\bf figure 3} by fixing $v_{rot}$ and varying $M_{A'}$ around 
the best fit
value, for a particular example. 
\vskip 1cm
\centerline{\epsfig{file=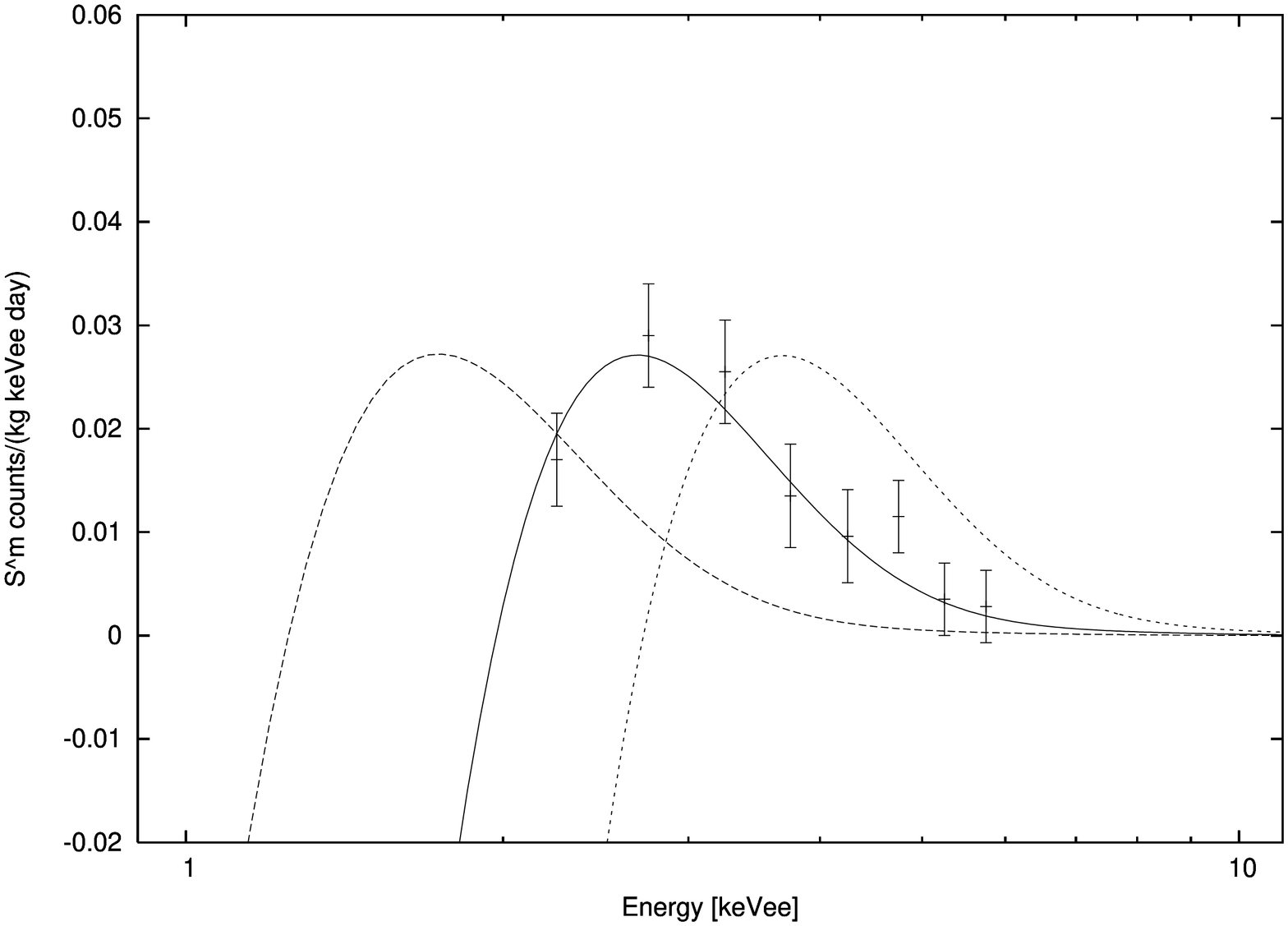,angle=270,width=13cm}}
\vskip 0.5cm
\noindent
Figure 3: Annual modulation amplitude for the case of $A' = O'$,
(solid line), $A' = C'$ (dashed line), $A' = Ne'$ (short-dashed line), ie. with $A' = 16
\pm 4$.
In each case we have used the same $v_{rot} = 280$ km/s, assumed a $He'$ mass 
dominated halo, and normalized the height of the peak to be the same.
This figure illustrates that the position of the peak determines $A'$, with
the shape of the energy spectrum being approximately unchanged.

\vskip 1cm

As figure 3 illustrates, the position of the peak provides a measurement of the mass
of the element $A'$. Allowing for the possible uncertainty in $v_{rot}$,
pins down the mass range for the element $A'$ to the range: 15 GeV $\stackrel{<}{\sim} 
M_{A'} \stackrel{<}{\sim}$ 30 GeV.  It is interesting that
this mass  range is theoretically 
consistent ($M_{He} < M_{A'} < M_{Fe}$) and also close to the naive expectation,
of oxygen mass. Oxygen is the most abundant heavy ordinary element in the Universe, and 
it would  not be  surprising  if the mirror sector  were  similar.

If we consider $A'$ as a parameter, then we can
evaluate the  $3\sigma$ ($5\sigma$) DAMA  allowed region
by finding  the  contours in ($v_{rot},  \ A'$) space where  
$\chi^2 = \chi^2_{min} + 9$ ($\chi^2 = \chi^2_{min}+25$),
where at each  point,  we vary $\epsilon \sqrt{\xi_{A'}}$ such that the normalization is correct.
We plot this  DAMA allowed region  in  {\bf figure 4}, for the case of a $He'$ mass dominated halo.
The case of a $H'$ dominated halo is very similiar.

\vskip 1cm

\centerline{\epsfig{file=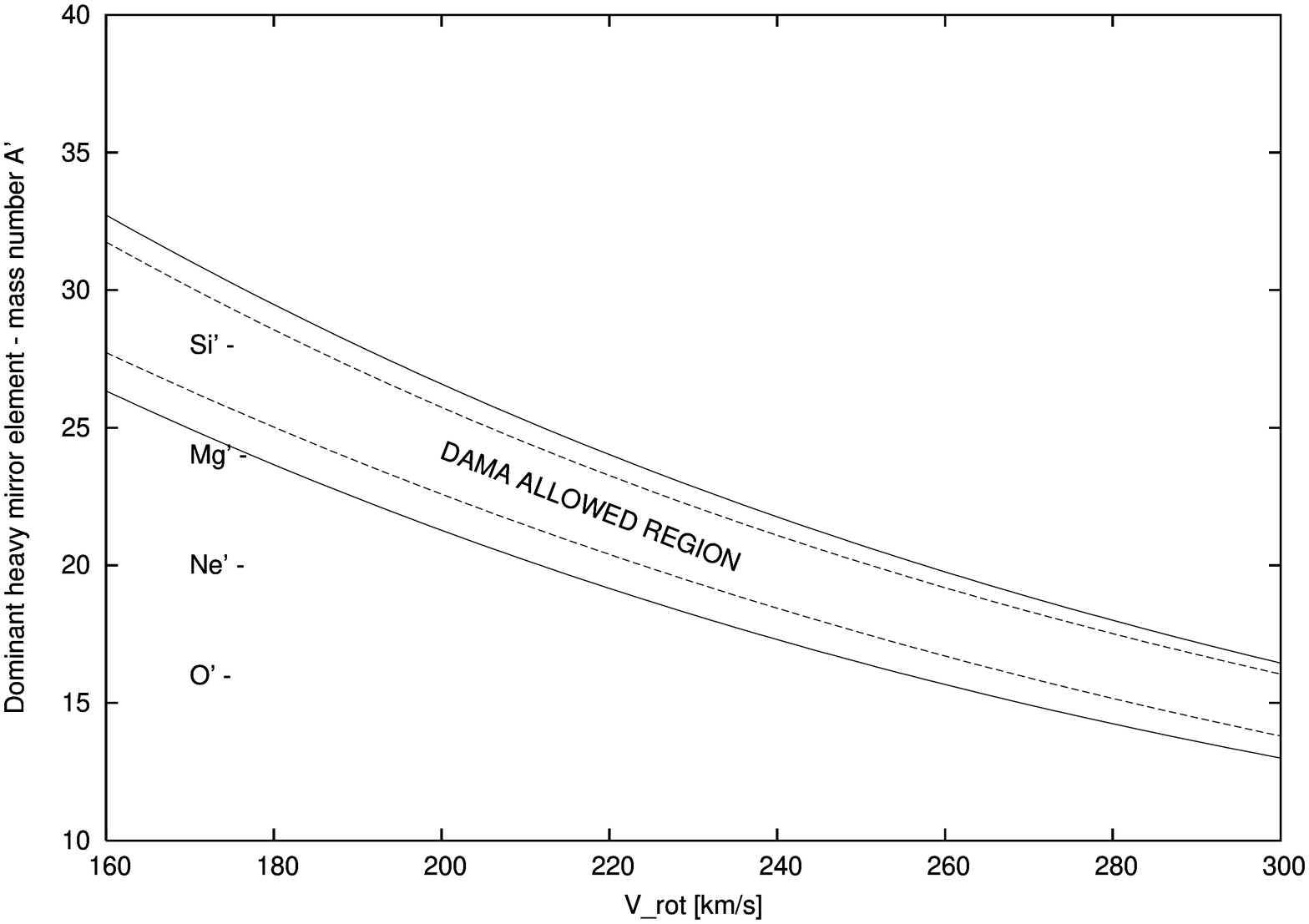,angle=270,width=11.5cm}}
\vskip 0.4cm
\noindent
Figure 4: The $3\sigma$ (dashed line) and $5\sigma$ (solid line) allowed region 
of $v_{rot}, A'$ parameter space 
consistent with the measured energy distribution of the modulation amplitude.
In each case we have fixed $\epsilon \sqrt{\xi_{A'}}$ 
so that the mean modulation amplitude is 0.0129 cpd/kg/keVee.

\vskip 0.5cm

The main features of our numerical  results can be approximately understood: 
The data shows evidence for a peak in the annual modulation  energy  spectrum. 
A peak is expected  
analytically.  As  we  discussed  earlier [Eq.(\ref{peak})], the position  of the peak
corresponds to  the  value of  $E_R$ (which we denote as $E_R^{peak}$) 
where $x \simeq y_0$,  or equivalently, to  the  value
of  $E_R$ where $v_{min} (E_R^{peak})\simeq v_{rot}$.  Evaluating  this expression, using
Eq.(\ref{v}), we  find:
\begin{eqnarray}
M_{A'} \simeq {M_A \over v_{rot}\sqrt{{2M_A \over  E_R^{peak}}} - 1}
\label{mon}
\end{eqnarray}
This shows again that
the value  of $E_R$ at the maximum  of  $dR^1/dE_R$, provides a 
determination  of  the mass of  the dominant heavy mirror  element $A'$.
The experimental data suggests that
the  maximum occurs roughly at $E_R \simeq 2.75$ keVee. Evaluating
Eq.(\ref{mon}) for $A = I$ and for  $A = Na$ gives:
\begin{eqnarray}
{M_{A'} \over M_p} &\simeq & {13.7 \over \left({v_{rot} \over 300 km/s}\right) - 0.107},
\ {\rm  for \ A = I}\nonumber \\
{M_{A'} \over M_p} &\simeq & {5.8 \over \left({v_{rot} \over  300  km/s}\right) - 0.25},
\ {\rm  for \ A =  Na}
\end{eqnarray}
Thus,  we see that  our  allowed region  corresponds to the case  where  $A'$ is  interacting
predominately with  $I$.  We  might have expected a second distinct  region of  allowed
parameter  space - a  second branch at  lower $v_{rot}$ -  
where the  rate  is dominated by interactions with  $Na$.
Numerically it turns  out that  the $Na$ peak is washed out  by  the  interactions of $I$,  whose
peak then occurs  around  $1.1$ keVee (if the peak in $Na$ is at $2.75$ keVee), 
and is  smeared  out into the  $E > 2$ keVee  range by the 
detector resolution. The  height of the $I$ peak is  also  enhanced  c.f.  with $Na$ peak, because
a) the basic cross-section on $I$ is larger than $Na$ because of the larger $I$ nuclear electric
charge,
b) the $I$ peak occurs  at 
lower $E_R$ and ${d\sigma \over  dE_R} \propto {1 \over  E_R^2}$, c) the  fraction
of channeled  events  strongly  increases towards  unity as  $E_R  \to 0$
in the adopted quenching  model  of ref.\cite{damaquench}.

Finally, in {\bf figure 5}, we give the absolute (unmodulated) spectrum, predicted by the model,
and compare it to the DAMA spectrum. Of course, the measured absolute spectrum has a significant
background component which is removed when
the time varying component is extracted. Figure 5 indicates that in this 
model the required background
spectrum is roughly consistent - it is smooth and positive. There is also
a hint of the large rise in absolute rate predicted by the model at low energies. Of course,
this is below the experimental threshold, and this hint should not be
taken too seriously.

Note that because of the sharply rising absolute spectrum at low $E_R$ our predicted shape 
of the absolute spectrum is somewhat sensitive to uncertainties in detector resolution
and quenching factor. The shape of the modulation spectrum, on the other hand,
is much less sensitive to these uncertainties.

\vskip 0.4cm
\centerline{\epsfig{file=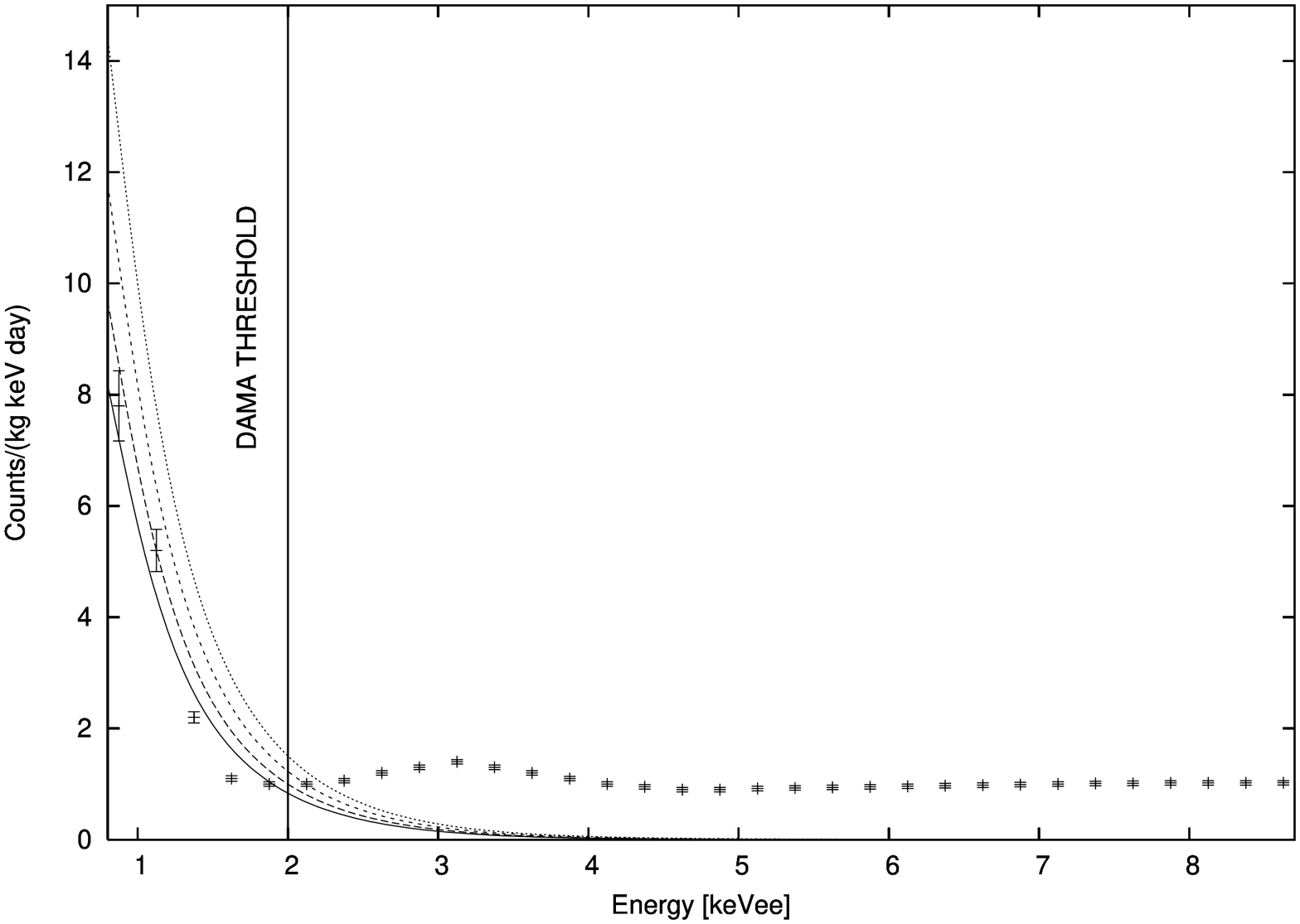,angle=270,width=11.4cm}}
\vskip 0.5cm
\noindent
Figure 5a:
Predicted absolute (unmodulated) rate, ${dR^0 \over dE_R}$, 
for the DAMA/NaI experiment. We take the same four examples of 
figure 2a: 
$A'=Si', v_{rot}=170$ km/s (solid line)
$A'=Mg', v_{rot}=195$ km/s (long-dashed line),
$A'=Ne', v_{rot}=230$ km/s (short-dashed line),
$A'=O', v_{rot}=280$ km/s (dotted line).
As with figure 2a, these results assume a $H'$ dominated halo by mass.
Also shown is the experimental (single hit) data.

\vskip 1cm
\centerline{\epsfig{file=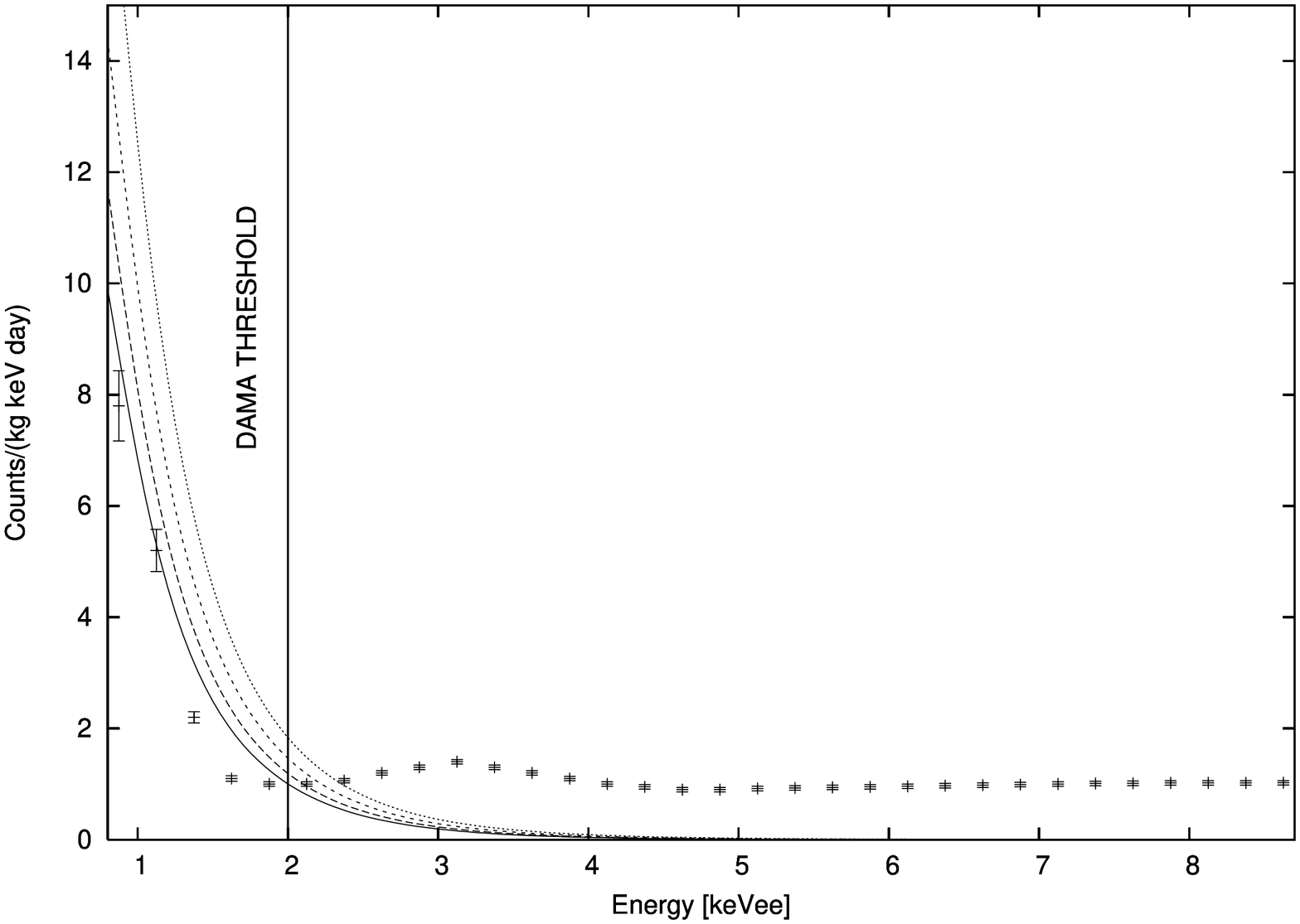,angle=270,width=11.4cm}}
\vskip 0.5cm
\noindent
Figure 5b: Same as figure 5a, except the mass of the halo is assumed to be dominated by $He'$.

\vskip 1cm

The value of $\epsilon \sqrt{\xi_{A'}}$ obtained from the fit to the DAMA/NaI and DAMA/LIBRA 
experiments suggests a value for $\epsilon$ in the range
$10^{-9}-10^{-8}$ for $10^{-3} < \xi_{A'} < 10^{-1}$.
Values of $\epsilon$ in this range have many interesting applications, see e.g.
ref.\cite{saibal,zurab}.
A mirror sector, interacting with the ordinary sector, with kinetic mixing
in this range is also consistent with Laboratory\cite{shelly} and
big bang nucleosynthesis constraints\cite{shelly2}.


\section{Constraints from the null experiments: CDMS/Ge, CDMS/Si, XENON10}

The annual modulation signal has only been seen in the DAMA/NaI and DAMA/LIBRA
experiments. Other direct detection experiments have not found
any dark matter signal. The most sensitive of these experiments
are the CDMS and XENON10 experiments.
These  experiments use sensitive 
background elimination techniques and  place limits on the  absolute dark matter event 
rate,  above their recoil energy detection  thresholds.

Importantly, these experiments are all higher threshold experiments, with claimed
nuclear recoil energy thresholds of 10 keV for CDMS/Ge, 7 keV for CDMS/Si and 4.5 keV for XENON10. 
The fact that mirror dark matter 
is light ($M_{A'} \sim M_{O}$)
has a narrow velocity dispersion, $v_O(A')^2 \ll v_{rot}^2$, and has cross-section proportional to
$1/E_R^2$ all help to enhance the signal in DAMA/NaI relative to these higher threshold
experiments. 

We have evaluated the contraints on our model from the CDMS/Ge\cite{cdms2}, CDMS/Si\cite{cdms1} 
and XENON10\cite{xenon}
experiments.  The published net 
effective exposures for these experiments are $121\ kg-d,\  12\ kg-d$
and  $\sim 100 \ kg-d$ for CDMS/Ge,  CDMS/Si  and XENON respectively.
As, indicated by figure 5, we expect a sharply falling absolute event rate.
Therefore the most sensitive energy region in these higher threshold experiments
will be near their energy threshold.
In the threshold region, CDMS/Ge\cite{cdms2}, CDMS/Si\cite{cdms1}
and XENON10\cite{xenon} found 0, 0 and 1 event respectively. The Poisson 
$95\%$ exclusion region is then $N=3$ for the CDMS
experiments and $N=4.74$ for the XENON experiment.
We have computed these exclusion limits for the parameter space of interest for the
DAMA experiment, which we show in {\bf figure 6}. As the  figure
clearly demonstrates,  our favoured  region of  parameter space is  consistent with the null results
of these experiments.

\vskip 0.8cm
\centerline{\epsfig{file=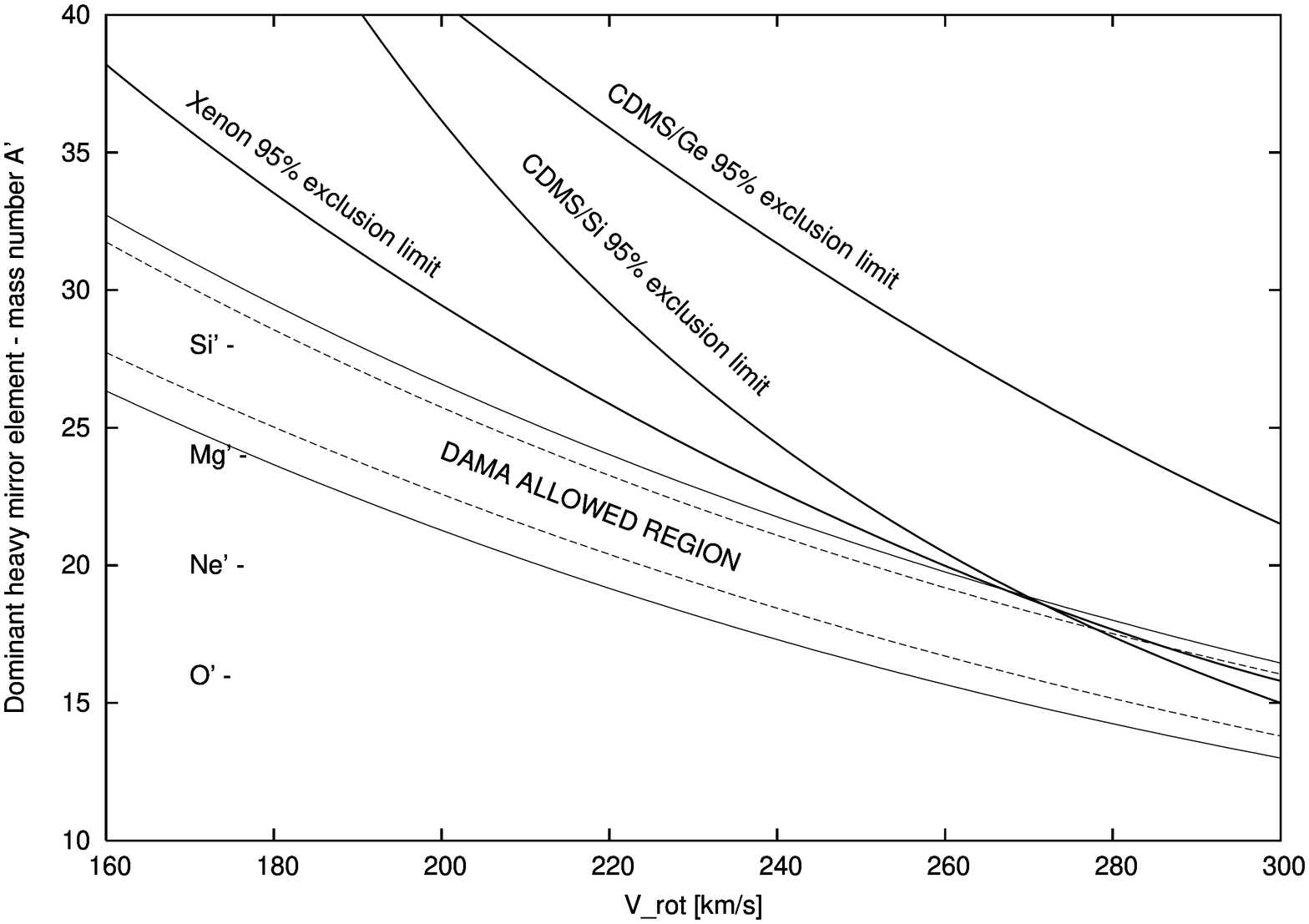,angle=270,width=13cm}}
\vskip 0.4cm
\noindent
Figure 6: DAMA allowed region, as shown in figure 4, except with the 
addition of the 95\% C.L. exclusion limits 
from the XENON10, CDMS/Ge and CDMS/Si experiments (the regions above the exclusion
contours  are the disfavoured region).

\vskip 1cm

Finally, note that there are a few experiments with lower thresholds than DAMA/NaI.
In particular there is the CRESSTI experiment with threshold of 0.6 keV\cite{cresst}.
The compatibility of the CRESST experiment with the 
DAMA/NaI annual modulation signal was analysed in the mirror
matter context in ref.\cite{f1}, but without taking into account
the channeling effect in the DAMA analysis. 
It was found that the predicted rate for the CRESST experiment was about a factor
of two higher than the measured rate. Including the channeling effect in the DAMA
analysis lowers the predicted cross section
by more than an order of magnitude, which means that the predicted
event rate for the CRESST experiment is around $10\%$ of their measured event rate. Thus the mirror dark
matter interpretation of the DAMA/NaI signal is completely consistent with the 
results of the low threshold CRESST experiment.  
Another low threshold experiment of note is the TEXONO experiment\cite{texono} with a claimed threshold
of 0.2 keV.  The low exposure and large background
currently limit the usefulness of that experiment. However, the sharply rising
event rate at low recoil energies, predicted by the mirror dark matter
explanation of the DAMA/NaI experiment, might potentially be probed by future
upgrades of experiments such as TEXONO and CRESST. The possibility of searching 
for annual modulation in such low threshold experiments would, of course, also be worthwhile. 

Finally note that there are earlier CDMS/Ge experiments with thresholds 
of 5 keV\cite{cdms5} and 7 keV\cite{cdms1}. However the lower exposure and high
backgrounds of these early runs make these experiments less sensitive than
the XENON10 experiment for the mirror dark matter interpretation of the DAMA experiment,
and therefore, within this theoretical framework, they do not disfavour any of the DAMA
allowed parameter space.

\section{Hidden sector dark matter models}

Mirror matter arises if one assumes an extra sector of particles and forces exactly isomorphic 
to the known particles. With the assumption that $T' \ll T$ in the early Universe,
this simple model is compatible with the inferred properties of dark matter, such
as its large scale structure, and, as we have shown,
with all of the direct detection experiments as well. 

Of course, one can imagine generic hidden sector models which can approximately mimic the 
effects of the  mirror
matter model for dark matter experiments. 
The simplest such model would have a) two stable particles with 
masses $M_1$ and $M_2$, such that $M_1 \ll M_2$ and b) interact with each other via
an unbroken $U(1)'$ (and possibly other gauge interactions), 
and c) interact with ordinary matter via $U(1)_Y \otimes U(1)'$
kinetic mixing.
With these assumptions, one can have $v_0 (M_2)^2 \ll v_{rot}^2$ if the mass density
of the halo is dominated by the lighter particles of mass $M_1$.  Provided 
that $15 \ {\rm GeV} \stackrel{<}{\sim} M_2 \stackrel{<}{\sim} 30$ GeV
such a model can give similar predictions to the mirror matter theory, and thus
is experimentally viable - at the moment.

However it is interesting that all these features automatically occur
in the mirror matter theory, which is theoretically tightly constrained. 
For example, in the mirror matter theory, $M_2$ is predicted
to be in the range
$M_{He} < M_{A'} \le M_{Fe}$. Thus, the mirror dark matter theory
is theoretically favoured over some generic alternative hidden sector model. 
This theoretical prejudice will be put to the test by future
experiments, which we await with interest.

\section{Conclusion}

Mirror dark matter arises  in simple and renormalizable extensions 
of the standard model whereby 
an exactly isomorphic  sector  of  particles and forces is hypothesized which  
interact with the known
particles via renormalizable $U(1)$ kinetic  mixing. Such extensions
of the standard model are theoretically well motivated, allowing
space-time parity to exist as an exact unbroken symmetry.
They also supply a spectrum of necessarily stable dark matter candidates of known
masses.

We have demonstrated that this dark matter theory can simply
and fully explain the annual modulation signal seen in the DAMA/NaI 
and DAMA/LIBRA experiments.
Our results indicate that the simplest and most plausible scenario: 
a predominant $H'/He'$  halo with
a small subdominant $\sim O'$ component is sufficient to explain the data. 
The required photon-mirror photon
mixing parameter is $\sim 10^{-9}$, which is consistent with 
all other experiments, and 
observations.
In particular, we
have shown that this simple scenario is also completely compatible with the latest null 
results of the other experiments, such as CDMS
and XENON10. 

Generic hidden sector models which approximately mimic the 
effects of mirror
matter type dark matter for direct detection experiments 
are also possible, but theoretically less appealing. 
Future experiments will, of course, be the final arbitrator.

\vskip 1cm
\noindent
{\bf Acknowledgements}
\vskip 0.4cm
\noindent
This work was supported by the Australian Research Council.

\end{document}